\documentstyle[12pt]{article}
 
\begin{document}
\def\omt{\tilde{\omega}}
\def\ti{\tilde}
\def\o{\Omega}
\def\bchi{\bar\chi^i}
\def\In{{\rm Int}}
\def\ba{\bar a}
\def\w{\wedge}
\def\ep{\epsilon}
\def\k{\kappa}
\def\Tr{{\rm Tr}}
\def\ST{{\rm STr}}
\def\ss{\subset}
\def\rn{\vert \alpha\vert^2}
\def\bi{\bibitem}
\def\ot{\oti\def\om{\omega}
\dmes}
\def\bc{{\bf C}}
\def\ptp{\stackrel{\otimes}{,}}
\def\br{{\bf R}}
\def\de{\delta}
 \def\bt{\beta}
 \def\ve{\vert}
\def\al{\alpha}
\def\la{\langle}
\def\ra{\rangle}
\def\G{\Gamma}
\def\st{\stackrel{\wedge}{,}}
\def\sto{\stackrel{\otimes}{,}}
\def\th{\theta}
\def\lm{\ti\lambda}
\def\U{\Upsilon}
\def\jp{{1\over 2}}
\def\js{{1\over 4}}
\def\d{\partial}
\def\tr{\triangleright}
\def\trl{\triangleleft}
\def\d{\partial}
\def\bq{\}_{P}}
\def\be{\begin{equation}}
\def\ee{\end{equation}}
\def\bea{\begin{eqnarray}}
\def\eea{\end{eqnarray}}
\def\D{{\cal D}}
\def\G{{\cal G}}
\def\K{{\cal K}}
\def\H{{\cal H}}
\def\R{{\cal R}}
\def\B{{\cal B}}
\def\T{{\cal T}}
\def\bT{\bar{\cal T}}
\def\F{{\cal F}}
\def\n{{1\over n}}
\def\si{\sigma}
\def\ta{\tau}
\def\ov{\over}
\def\l{\lambda}
\def\L{\Lambda}

\def\pih{\hat{\pi}}
\def\Vt{V^{\ti}}
\def\Ut{U^{\ti}}
\def\e{\varepsilon}
\def\bt{\beta}
\def\ga{\gamma}
\def\om{\omega}
\def\be{\begin{equation}}
\def\ee{\end{equation}}
\def\bea{\begin{eqnarray}}
\def\eea{\end{eqnarray}}
\def\D{{\cal D}}
\def\G{{\cal G}}
\def\H{{\cal H}}
\def\R{{\cal R}}
\def\B{{\cal B}}
\def\K{{\cal K}}
\def\T{{\cal T}}
\def\bT{\bar{\cal T}}
\def\F{{\cal F}}
\def\n{{1\over n}}
\def\si{\sigma}
\def\ta{\tau}
\def\ot{\otimes}
\def\l{\lambda}
\def\L{\Lambda}
\def\ve{\vert}

\def\pih{\hat{\pi}}
\def\Vt{V^{\ti}}
\def\Ut{U^{\ti}}
\def\e{\varepsilon}
\def\bt{\beta}
\def\ga{\gamma}

\sloppy \raggedbottom
\setcounter{page}{1}

\newpage
\setcounter{figure}{0}
\setcounter{equation}{0}
\setcounter{footnote}{0}
\setcounter{table}{0}
\setcounter{section}{0}

\begin{titlepage}
\begin{flushright}
{}~
IML 2005-22\\
hep-th/0511033
\end{flushright}

\vspace{3cm}
\begin{center}
{\Large \bf  Poisson-Lie symmetry and $q$-WZW model}\\ 
[50pt]{\small
{\bf C. Klim\v{c}\'{\i}k }
\\ ~~\\Institute de math\'ematiques de Luminy,
 \\163, Avenue de Luminy, 13288 Marseille, France}

\vspace{1cm}

\begin{abstract}
We review    the notion of (anomalous) Poisson-Lie symmetry of a dynamical system  and we outline 
the  Poisson-Lie symmetric deformation of  the standard WZW model from the vantage point of the
twisted Heisenberg double.  \end{abstract}
\vspace{1cm}
\noindent {\it  To appear in the Proceedings of the 4th International  Symposium On Quantum Theory And Symmetries (QTS-4), Varna Free University,  Bulgaria, 15-21 August 2005}

\end{center}
\end{titlepage}
\newpage

\section{Introduction}
The concept of Poisson-Lie symmetry  \cite{ST1} of a dynamical system  is a generalization of    ordinary Hamiltonian symmetry and it is characterized by the presence of two Lie groups in the story. One of those groups  is called a symmetry group
and it acts on a phase space of a dynamical system. The structure of the other group (called  cosymmetry group) underlies the way how the action of the symmetry group is expressed in terms of Poisson brackets. If the cosymmetry group is Abelian the Poisson-Lie symmetry is nothing but  the
ordinary Hamiltonian symmetry known from every textbook on classical mechanics.   Sometimes
the Poisson-Lie symmetry can be "switched off" and it becomes the ordinary symmetry. This happens
for one-parameter families of Poisson-Lie symmetric dynamical systems such that for a particular
value of the parameter  the cosymmetry group becomes Abelian (see Example in Section 3). In such
case we say that  the corresponding ordinary symmetry is deformable.  This short contribution
has two purposes. First of all it aims to be a very concise review of the concept of the (anomalous) Poisson-Lie symmetry and, secondly, it wants to present the main results of the   paper \cite{K} from the vantage point of the so-called twisted Heisenberg double  \cite{ST}.
\vskip1pc
\noindent In Sec 2, we review the definition of the Poisson-Lie group and  in Sec 3, we describe and illustrate  the concept of the Poisson-Lie symmetry. In Sec 4, we review the construction of the twisted Heisenberg double due to  Semenov-Tian-Shansky and we show that the structure of the standard WZW model can be understood in its terms.  Finally, in Sec 5, we show how to choose the twisted Heisenberg double in order to obtain the $q$-deformation of the WZW model.

\section{Poisson-Lie groups}

Let $B$ be a Lie group and $Fun(B)$ the algebra of   functions on it.  It is well known that the group structure
on $B$ can be (dually) described by the  so called coproduct $\Delta:Fun(B)\to  Fun(B)\otimes Fun(B)$, the antipode
$S:Fun(B)\to Fun(B)$ and the counit $\e:Fun(B)\to \br$ given, respectively,  by the formulae
$$\Delta y(b_1,b_2)=y'(b_1)y''(b_2)=y(b_1b_2), \quad S(y)(b)=y(b^{-1}), \quad  \e(y)=y(e_B).$$
Here $y\in Fun(B)$, $b,b_1,b_2\in B$,  $e_B$ is the unit element of $B$ and we use the Sweedler notation for the coproduct: $$\Delta y= \sum_\al y'_\al\otimes y''_\al\equiv y'\otimes y''.$$ 
Poisson-Lie group is a Lie group equipped with a Poisson-bracket $\{.,.\}_B$ such that
$$\Delta\{y_1,y_2\}_B=\{y_1',y_2'\}_B\otimes y_1''y_2''+y_1'y_2'\otimes \{y_1'',y_2''\}_B, \quad y_1,y_2\in Fun(B).\eqno(1)$$
Consider the linear dual $\B^*$ of the Lie algebra $\B=Lie(B)$. The Poisson-Lie bracket $\{.,.\}_B$ induces a natural Lie algebra structure $[.,.]^*$ on $\B^*$.   Let us explain this fact in more detail: 
First of all recall that $\B^*$ can be identified with the space of right-invariant $1$-forms on the group
manifold $B$.  We have a natural (surjective)  map   $\phi: Fun(B)\to \B^*$   defined by 
$$\phi(y)= dy'S(y''), \quad y\in Fun(B).$$
Note that the $1$-form $\phi(y)$ is   right-invariant  therefore it is indeed in $\B^*$.  Let 
$u_1,u_2\in \B^*$ and $y_1,y_2\in Fun(B)$ such that $u_j=\phi(y_j), j=1,2$. Then we have
$$[u_1,u_2]^*=\phi(\{y_1,y_2\}_B).$$
It is the Poisson-Lie property (1) of  $\{.,.\}_B$ which ensures the independence of $[u,v]^*$ on the
choice of the representatives $y_1,y_2$. In what follows, the Lie algebra $(\B^*, [.,.]^*)$ will be denoted 
by the symbol $\G$  and $G$ will be a (connected simply connected) Lie group such that $\G=Lie(G)$.
We note that $G$ is often referred to as the dual group of $B$.

\section{Poisson-Lie symmetry}

A dynamical system is a triple $(M,\om,v)$ where $M$ is a   manifold,
$\om$ is a non-degenerate closed 2-form on it and $v$ is a vector field on $M$ defining the time evolution. One often
considers the case when the  vector field $v$ is Hamiltonian. This means that it exists a function $H\in Fun(M)$ such that
$$v=\Pi_M(.,dH).$$
Here $\Pi_M$ is the so-called Poisson bivector (i.e. an antisymmetric contravariant tensor field) on $M$ obtained by the inversion of the symplectic form $\omega$. An expression $\Pi_M(df,dg)$  for $f,g\in Fun(M)$ is nothing but
  the Poisson bracket  $\{f,g\}_M$. 
\vskip1pc

\noindent \underline{Basic definition}: 
 \vskip1pc
\noindent Let $B$ be a Poisson-Lie group and $G$ its dual group.   We call the dynamical system $(M,\om,v)$ $(G,B)$-Poisson-Lie symmetric if $G$ acts on $M$ and if it exists a surjection
$\mu:M\to B$  (called the momentum map)  such that
\vskip3pt
\noindent i) The image $Im(\mu^*)$ of the (dual) momentum map is the Poisson subalgebra of $Fun(M)$ stable with respect to the evolution vector field $v$.

\noindent ii)  The map $w:Fun(B) \to Vect(M)$ given by
$$w(y)\equiv \Pi_M(.,\mu^*\phi(y)), \quad y\in Fun(B)$$
is the homomorphism of Lie algebras fulfilling $Im(w)=Lie(G)$.

\vskip1pc
\noindent \underline{Remarks and explanations}:
 \vskip1pc
\noindent  a)  The group
$G$ is called the symmetry group since it is the one which acts on the phase space $M$.
$B$ is  called the cosymmetry group and it underlies (via $\phi$)  the way how the
$G$-action is described via the Poisson bracket on $M$.
 
\noindent b) Both $Fun(B)$ and $Vect(M)$ are naturally Lie algebras, the former with respect to the Poisson bracket $\{.,.\}_B$, the latter with respect to the Lie bracket of vector fields on $M$. When writing $Im(w)=Lie(G)$ we mean that, using the action of $G$ on $M$,  $Lie( G)$ is embedded in $Vect(M)$.

\noindent  c) If the dual   map $\mu^*:Fun(B)\to Fun(M)$ is a Poisson morphism 
(i.e. if $\{\mu^*y_1,\mu^*y_2\}_M=\mu^*\{y_1,y_2\}_B$) the Poisson-Lie symmetry is called non-anomalous. If $\mu^*$ is not the Poisson morphism the symmetry is called anomalous.
 
\vskip1pc
 
\noindent \underline{Example of non-anomalous Poisson-Lie symmetry}:
\vskip1pc 
 
\noindent  For $G$ we take the group $SU(2)$ whose elements are matrices of the form
  $\left(\begin{array}{cc} \al &-\bar\bt\\ \bt  &\bar \al\end{array}\right)$, $\al,\bt\in\bc,$, $\al\bar\al+\bt\bar\bt=1$.
  For $B$ we take a three-dimensional manifold $\br^3$ with the usual cartesian coordinates
denoted as $J^i\in Fun(B), i=1,2,3$. The group structure  on $B$ is (dually) defined by the coproduct, the antipode and  the counit 
$$\Delta J^1=e^{\ep J^3}\ot J^1 +J^1\ot e^{-\ep J^3}, \eqno(2a)$$
$$\Delta J^2=e^{\ep J^3}\ot J^2 +J^2\ot e^{-\ep J^3}, \eqno(2b)$$
$$\Delta J^3=1\ot J^3 +J^3\ot 1, \eqno(2c)$$
$$S(J^i)=-J^i,\quad \e(J^i)=0, \quad i=1,2,3.\eqno(2d)$$
The Poisson-Lie bracket on $B$ is given  by
 $$\{J^3,J^1\}_B=J^2,\quad \{J^3,J^2\}_B=-J^1,\quad \{J^1,J^2\}_B={{\rm sinh}(2\ep J^3)\over 2\ep}.\eqno(3) $$
Let us now describe a $(G,B)$-dynamical system $(M,\om_\ep,v)$. For $M$ we take $\bc^2(=\br^4)$
and we parametrize it by two complex coordinates $A,B\in Fun^{\bc}(M)$ given in terms of the 
cartesian coordinates of $\br^4$ as  $A=x^1+ix^2$, $B=x^3+ix^4$. For the  ($\ep$-independent)
evolution vector field $v$ we take 
$$v=-{i\over 2}(A\bar A+B\bar B)\biggl(\bar A{\d \over \d {\bar A}}+ \bar B{\d \over \d {\bar B}}- A{\d \over \d {A}}- B{\d \over \d { B}}\biggr)$$
Instead of detailing the 
symplectic form $\om_\ep$, we  directly describe the Poisson bracket which it induces. Thus:
$$\{A,\bar A\}_M=i\sqrt{1+\ep^2(A\bar A+B\bar B)^2}+i\ep B\bar B,$$
$$\{B,\bar B\}_M=i\sqrt{1+\ep^2(A\bar A+B\bar B)^2}-i\ep A\bar A,$$
$$\{A,B\}=-i\ep AB, \quad \{\bar A,\bar B\}=i\ep \bar A\bar B,\quad  \{A,\bar B\}=0, \quad \{\bar A,B\}=0.$$
Note that in the limit $\ep\to 0$ the bracket on $M=\br^4$ becomes just the Darboux Poisson bracket.
\vskip1pc
\noindent The action of $G=SU(2)$ on $M$ is given by
 
$$\left(\begin{array}{cc} A\\ B \end{array}\right)\to   \left(\begin{array}{cc} \al &-\bar\bt\\ \bt &\bar \al   \end{array}\right)\left(\begin{array}{cc} A\\ B \end{array}\right)\eqno(4)$$
The comomentum map $\mu_\ep^*$ reads
$$\mu_\ep^*(J^3)= -{1\over 2\ep}{\rm ln}\biggl (\sqrt{1+\ep^2(A\bar A+B\bar B)^2}+\ep(B\bar B-A\bar A)\biggr),$$
$$\mu_\ep^*(J^1)=\jp (\bar A B +A\bar B)\exp{(\ep \mu_\ep^*(J^3))},$$
$$\mu_\ep^*(J^2)={i\over 2} (\bar AB -A\bar B)\exp{(\ep \mu_\ep^*(J^3))},$$
It is straightforward to verify that $Im(\mu^*_\ep)$ is  the Poisson subalgebra of $Fun(M)$. Indeed, we have 
 $$\{J^3,J^1\}_M=J^2,\quad \{J^3,J^2\}_M=-J^1,\quad \{J^1,J^2\}_M={{\rm sinh}(2\ep J^3)\over 2\ep}.$$
where we have (somewhat abusively)  written   just $J^j$ instead of $\mu^*_\ep(J^j)$. It turns out that the evolution vector  field $v$ is Hamiltonian. The 
Hamiltonian $H_{\ep}$   reads
$$H_\ep= (J^1)^2+(J^2)^2+ {{\rm sinh}^2(\ep J^3)\over \ep^2}$$
and fulfils
$$\{H,J^k\}_M=0, \quad k=1,2,3.$$
Thus $Im(\mu^*)$ is indeed evolution invariant as it should.  
\vskip1pc
\noindent Now we calculate
$$\phi(J^1)=e^{\ep J^3} dJ^1 -  J^1 d(e^{\ep J^3}), $$
$$\phi(J^2) =e^{\ep J^3} dJ^2 -  J^2 d(e^{\ep J^3}), $$
$$\phi(J^3)=dJ^3.$$
The map $w$ is indeed the Lie algebra homomorphism (this can be checked by using (3) ) and it  gives three vector fields acting on $f\in Fun(M)$:
$$w(J^1)f=e^{\ep J^3}\{f,J^1\}_M- J^1 \{f,e^{\ep J^3}\}_M,$$
$$w(J^2)f=e^{\ep J^3}\{f, J^2\}_M- J^2  \{f,e^{\ep J^3}\}_M,$$
$$w(J^3)f=\{f,J^3\}_M.$$
Whatever is the value of $\ep$, the vector fields $w(J^k)$ turn out to be   the same:
$$w(J^1)=-{i\over 2}\biggl(\bar B{\d \over \d {\bar A}}+ \bar A{\d \over \d {\bar B}}- B{\d \over \d {A}}- A{\d \over \d { B}}\biggr)$$
$$w(J^2)=-{1\over 2}\biggl(\bar B{\d \over \d {\bar A}}- \bar A{\d \over \d {\bar B}}+B{\d \over \d {A}}- A{\d \over \d { B}}\biggr)$$
$$w(J^3)=-{i\over 2}\biggl(\bar A{\d \over \d {\bar A}}-\bar B{\d \over \d {\bar B}}- A{\d \over \d {A}}+B{\d \over \d { B}}\biggr)$$
They verify
$$ [w(J^j),w(J^k)]= -\ep^{jkl}w(J^l),$$
where $\ep^{jkl}$ is the well-known alternating symbol. 
For completeness, let us mention that the vector fields $w(J^k)$  indeed encode the infinitesimal version of the action (4) of the group $G=SU(2)$ on $M=\bc^2$, as they should.

\vskip1pc
\noindent \underline{Deformation program}
\vskip1pc

\noindent Let us first  clarify the role of the (real) parameter $\ep$ in the example above.  Actually, Eqs. (2abcd)   define  a one-parameter family $B_\ep$ of the cosymmetry groups. Note that for   $\ep=0$  the group $B_{0}$ is Abelian (the group law   becomes  just the addition of vectors in $\br^3$) and the  brackets (3)  become the Kirillov-Kostant  brackets on the dual of the Lie algebra $su(2)(=\br^3)$.  Thus, for $\ep=0$, the vector fields $w(J^k)$ become all Hamiltonian (i.e. $w(J^k)f= \{f,J^k\}_M$) and the dynamical system $(M,\om_0,v)$  is symmetric in the ordinary sense.  
\vskip1pc
\noindent More generally, having a dynamical system $(M,\om,v)$ possessing an ordinary symmetry with respect to the group $G$ acting on $M$ in the Hamiltonian way, the deformation program consists in finding a one-parameter family of cosymmetry groups $B_\ep$ and a one parameter family of
$(G,B_\ep)$- Poisson-Lie symmetric dynamical systems $(M,\om_\ep,v)$ in such a way that for $\ep=0$
we recover the original system $(M,\om,v)$.  In particular, this means that $\om=\om_0$ and  $B_0$ becomes the Abelian group.   
\vskip1pc
\noindent Note that in the deformation process the manifold $M=\bc^2$ was not deformed, neither the action of the symmetry group $G=SU(2)$ on $M$ and not even the evolution vector field $v$. Only the symplectic form $\om$ and the cosymmetry group $B$ got deformed.  Such situation takes place also for $q$-WZW
model.

\section{Twisted Heisenberg double}

The basic observation which triggered our work is as follows: The ordinary WZW model is a particular example of a  general construction
 of (anomalous) Poisson-Lie symmetric systems known under the name of twisted Heisenberg double \cite{ST}. Let us describe this construction in more detail by stating two lemmas:
\vskip1pc
 
\noindent \underline{Lemma 1}:
 \vskip1pc
\noindent  Consider a  Lie group $D$ equipped with a non-degenerated bi-invariant  metric and possessing two maximally isotropic  subgroups $G$ and $B$. Let $\k$ be a  metric preserving 
automorphism of $D$   and let $T^i$ and $t_i$ be the respective basis of $\G=Lie(G)$ and $\B=Lie(B)$ such that  $(T^i,t_j)_\D=\delta^i_j$. Then the (basis independent) expression 
$$\{f_1,f_2\}_D\equiv \nabla^R_{T^i}f_1 \nabla^R_{t_i}f_2 -\nabla^L_{\k(t_i)}f_1\nabla^L_{\k(T^i)}f_2, \quad f_1,f_2\in Fun(D) \eqno(5) $$
is a non-degenerated Poisson bracket making $D$ a symplectic manifold.  \vskip1pc

\noindent \underline{Remarks and explanations}:
\vskip1pc
\noindent a) The group $D$ equipped with the Poisson bracket $\{.,.\}_D$ is called the twisted  Heisenberg double of $G$. We denote as $\om_D$  the symplectic form corresponding to 
$\{.,.\}_D$.

\noindent b)  Bi-invariant  means both left-  and right-invariant. The  non-degenerated 
bi-invariant metric on $D$
obviously  induces an $Ad$-invariant non-degenerate bilinear form  $(.,.)_{\D}$ on $\D=Lie(D)$.
 An  isotropic submanifold  of $D$ is such that the induced metric on it vanishes. Maximally isotropic
 means  that it is not contained in any bigger isotropic submanifold.

\noindent c) The   vector fields $\nabla^{L,R}_T$ are defined as  
$$\nabla^L_T f(K)=\biggl({d\over ds}\biggr)_{s=0}f(e^{sT}K), \quad \nabla^R_T f(K)=\biggl({d\over ds}\biggr)_{s=0}f(Ke^{sT}),$$
where  $K\in D,T\in \D.$
\vskip1pc
\noindent \underline{Lemma 2}:
\vskip1pc
\noindent Suppose that the automorphism $\k$ preserves the subgroup $B$ and  two global unambigous decompositions holds: $D=BG$ and $D=\k(G)B$.  Consider (surjective) maps $\L_{L}, \L_R: D\to B$ induced by the decompositions $D=BG$  and $D=\k(G)B$, respectively. Let $v$ be any vector field leaving invariant the images 
of their dual maps   $\L^*_{L},\L^*_R: Fun(B)\to Fun(D)$.   Then $(D,\om_D,v)$ is the  Poisson-Lie symmetric
dynamical system with the symmetry group $G\times G$ and the cosymmetry group $B\times B$.
The symplectic form $\omega_D$ reads
$$\om_D=\jp(\L^*_L\rho_B\st \rho_D)_\D - \jp(\L^*_R\rho_B\st \l_D)_\D,\eqno(6)$$ the group $G\times G$ acts as 
$$(h_L,h_R)\tr l=\k(h_L)lh_R^{-1}, \quad h_L,h_R\in G,\quad  l\in D$$
and the momentum map $\mu:D\to B\times B$  is given by 
$\mu=\Lambda^L\times \Lambda^R.$
\vskip1pc

\noindent \underline{Remarks and explanations}:
\vskip1pc
\noindent   a)  Global unambigous decomposition $D=\k(G)B$ means that for every element $l\in D$ it exists a unique $g\in G$ and a unique $b\in B$ such that $l=\k(g)b^{-1}$. Similarly for $D=BG$:
it exists a unique $\ti g\in G$ and a unique $\ti b\in B$ such that  $l=\ti b\ti g^{-1}$.

\noindent  b) $\rho_D$ ( $\l_D$) is right(left)-invariant Maurer-Cartan form on the group $D$,
$\L^*_{L(R)}\rho_B$ is the pull-back of the right-invariant Maurer-Cartan form on $B$ by the   map $\L^*_{L(R)}$.

\vskip1pc

\noindent \underline{Example: The standard WZW model}.
 \vskip1pc
 
\noindent  
 Let $K$ be a simple connected and simply connected compact Lie group and $LK$ the 
group of smooth maps from a circle $S^1$ into $K$ (the group law is given by pointwise
multiplication). It is important for us that there  exists a natural non-degenerate invariant bilinear form $(.\vert .)$ on $ L\K \equiv  Lie (LK)$ given by formula
$$(\al\ve \bt)={1\over 2\pi}\int d\si (\al(\si),\bt(\si))_{\K},   \eqno(7)$$
where the angle variable $\si\in [-\pi,\pi]$ parametrizes the circle $S^1$ and $(.,.)_{\K}$ is the (appropriately normalized) Killing-Cartan form on $\K\equiv Lie(K)$. 
\vskip1pc
\noindent For the twisted Heisenberg double $D$ we take the semidirect product
of $LK$ with its Lie algebra $L\K$.  Thus the group multiplication law on $D$  reads
$$(g,\chi).(\ti g,\ti\chi)=(g\ti g,\chi +Ad_g\ti\chi), \quad g\in LK, \chi\in L\K.$$
   Lie algebra $\D$ of $D$ has the structure of semidirect 
 sum $\D =L\K\stackrel{\to}{ \oplus} L\K$ where the second composant of the semidirect sum has trivial zero bracket. Thus
 $$[\chi\oplus \al,\xi\oplus \bt] =[\chi,\xi]\oplus ([\chi,\bt]-[\xi,\al]),$$
 where $\chi,\xi\in L\K$ are in the first and $\al,\bt\in L\K$ in the second composant of the semidirect sum. The bi-invariant metric on $D$ comes from $Ad$-invariant
 bilinear form $(.,.)_{\D}$ on $\D$ defined with the help of (7):
 $$(\chi\oplus\al,\xi\oplus\bt)_{\D}=(\chi\ve \bt)+(\xi\ve \al). $$
  The  metric preserving automorphism $\k$ of  the group $D$ reads
 $$\k(g,\chi)= (g,\chi +k\d_{\si} gg^{-1})$$
 where $k$ is an (integer) parameter.  The maximally isotropic subgroups are $G=LK$ and
 $B=L\K$. Note that $B$ is Abelian since the group law is given just by the addition of vectors in $L\K$.
  It is simple to establish the decompositions $D=\k(G)B$ and $D=BG$. Indeed, we have for every
  $g\in LK,\chi\in L\K$
 $$(g,\chi) = (g,k\d_\si gg^{-1})(e,Ad_{g^{-1}}\chi-kg^{-1}\d_\si g)=(e,\chi).(g,0), $$
 where $e$ is the unit element  of $LK$. 
\vskip1pc
\noindent The symplectic form $\om_D$ on $D$  is given by (6) and it  reads (cf. \cite{Gaw,Mad})
$$\om_{D}=d(J_L\ve  \rho)+\jp k(\rho {\ve} \d_{\si}\rho),$$
where  $\rho$ is  a $L\K$-valued  right-invariant  Maurer-Cartan form on $LK$ often written as $\rho =dgg^{-1}$ and $J_L$ is $L\K$-valued function on $D$ defined by
$$J_L(g,\chi)=\chi.$$In order to define the Hamiltonian $H$ we need one more $L\K$-valued function
on $D$ denoted $J_R$. Thus
$$J_R(g,\chi) = -Ad_{g^{-1}}\chi  +kg^{-1}\d_\si g$$ and
$$H=-{1\over 2k}(J_L\vert J_L) -{1\over 2k}(J_R\vert J_R).$$
 Let us study  the symmetry structure of our dynamical system $(D,\om_D,H)$.  First of all, the symmetry group   $G\times G=LK\times LK$ acts on $D$  as follows
 $$(h_L,h_R)\tr(g,\chi)= (h_L gh_R^{-1},k\d_\si h_L h_L^{-1}+h_L\chi h_L^{-1}), \quad h_L,h_R,g\in LK,\chi\in L\K.\eqno(8)$$
 The cosymmetry group $B\times B$ is   $L\K \oplus L\K$  (with Abelian group law given by the   addition of vectors). The (commutative) algebra $Fun(L\K)$ is generated by linear functions $F_{\chi}$ of the
 form
 $$F_{\chi}(\xi)= (\chi\ve \xi), \quad \chi,\xi\in L\K,$$
 hence the algebra $Fun(B\times B)=Fun(L\K)\otimes Fun(L\K)$ is generated by linear functions $F^{L}_\chi,F^{R}_\chi$. The Abelian
 group law on $B\times B$ is then encoded in the coproduct, antipode and counit on $Fun(B\times B)$:  
 $$\Delta F^{L(R)}_{\chi}=1\otimes F^{(L)R}_{\chi} +F^{(L)R}_{\chi}\otimes 1, \quad S(F^{L(R)}_\chi)=F^{L(R)}_{-\chi}, \quad \e(F^{L(R)}_{\chi})=0. $$
 The Poisson-Lie bracket on $B\times B$ is given by
  $$\{F_\chi\ot 1,F_\xi\ot 1\}_{B\times B}= F_{[\chi,\xi]}\ot 1, \eqno(9a)$$
 $$\{1\ot F_\chi,1\ot F_\xi\}_{B\times B}=1\ot F_{[\chi,\xi]},\eqno(9a)$$ 
  $$\{F_\chi\ot 1,1\ot F_\xi\}_{B\times B}=0,\eqno(9c)$$
  The (dual) momentum
 map $\mu^*:Fun(B\times B)\to Fun(D)$ is simply
 $$\mu^*(F^L_\chi) =\mu^*(F_{\chi}\otimes 1)  = (J_L\ve \chi) , \quad \mu^*(F^R_\chi) =\mu^*(1\otimes F_{\chi})=(J_R\ve \chi), \quad \chi\in L\K.$$
 In order to illustrate  that the WZW model $(D,\omega_{\D},H)$ is indeed $(G\times G,B\times B)$-Poisson-Lie symmetric, we have to verify
 the items i) and ii) of the \underline{Basic definition} of Section 3. First we notice that $Im(\mu^*)$ is the Poisson subalgebra of $Fun(D)$. Indeed, it is not difficult to calculate 
 $$\{(J_L\ve \chi), (J_L\ve \xi) \}_D=(J_L\ve ([\chi,\xi]) + k(\chi\ve \d_\si \xi), \eqno(10a)$$
 $$\{(J_R\ve \chi), (J_R\ve \xi) \}_D=(J_R\ve ([\chi,\xi]) - k(\chi\ve \d_\si \xi), \eqno(10b)$$ 
 $$\{(J_L\ve \chi), (J_R\ve \xi) \}_D=0.\eqno(10c)$$
 (By comparing (9abc) with (10abc) we see that $\mu^*$ is not the Poisson morphism hence the
 symmetry is anomalous). 
 Then we verify, that $Im(\mu^*)$ is stable with respect to the time evolution. Indeed, we have
 $$\{(J_L\ve \chi),H\}_D= (J_L\ve \d_\si \chi).$$
  $$\{(J_R\ve \chi),H\}_D= -(J_R\ve \d_\si\chi).$$
 Verifying the  item ii) of the basic definition is also easy. First we  identify $Fun(D)=Fun(LK)\otimes Fun(L\G)$ and, by inverting the symplectic form $\om_D$, we find 
  $$\{f_1\otimes 1,f_2\otimes 1\}_D=0, \quad \{f\otimes 1, 1\otimes  F_\chi\}_D=\nabla^L_\chi f\ot 1,$$
 $$\{1\ot F_\chi,1\ot F_\xi\}_D=1\ot F_{[\chi,\xi]} + k(\chi\ve \d_\si \xi) 1\ot 1.$$
 Then it is easy to calculate
  $$w((J_L\ve \chi))(f\otimes 1)=\{f\otimes 1, (J_L\ve \chi)\}_M=(\nabla^L_\chi f\otimes 1), \quad f\in Fun(LK).$$
  $$w((J_L\ve \chi))(1\otimes F_\xi)=\{1\otimes F_\xi, (J_L\ve \chi)\}_D=1\otimes F_{[\xi,\chi]}+k(\xi\ve \d_\si\chi)1\otimes 1,  $$
$$w((J_R\ve \chi))(f\otimes 1)=\{f\otimes 1, (J_R\ve \chi)\}_D=-(\nabla^R_\chi f\otimes 1), \quad f\in Fun(LK).$$
 $$w((J_R\ve \chi))(1\otimes F_\xi)=0.$$
 We thus see that the vector fields  $w((J_L\ve \chi))$  and  $w((J_R\ve \chi))$ indeed generate the infinitesimal
 version of the action (8) of the symmetry group $LK\times LK$ on the algebra of observables $Fun(D)$.
 
 \section{$q$-WZW model}
The deformation program outlined in Section 3 can be applied to
   the standard WZW model (cf. \cite{K,K1}).  The resulting theory is called a quasitriangular
   WZW model and it can be described in rather explicit fashion by performing the so-called chiral
   decomposition. The technical presentation of the construction of the $q$-WZW is however lengthy and it cannot be presented on a small space of few pages. Therefore we shall restricts ourselves to the very beginning and the very end of the story:
   
   \vskip1pc
   \noindent The twisted Heisenberg double underlying the $q$-WZW model
 is the loop group 
$LK^\bc$ consisting
of smooth maps from the circle $S^1$ into the complexified group $K^\bc$.
 (For example, the complexification of $SU(2)$ is $SL(2,\bc)$). It is important to stress
 that $LK^\bc$ is viewed as \underline{real} group.
 The invariant
nondegenerate bilinear form   $(.,.)_{\D}$ on $\D=Lie({LK^\bc})$ is then defined as 
$$ (x,y)_\D={1\over \ep} Im (x\vert y), \quad x,y\in \D,$$ 
 where $(.\ve .)$ is just the bilinear form (7) naturally extended to $L\K^\bc$ , $Im$ stands for the imaginary part (not for the image of a map as before!) and $\ep$ is the deformation parameter.  The metric preserving automorphism $\k$ of 
 $LK^\bc$ is defined most easily if we view the group $LK^\bc$ as a   group of holomorphic
maps from a Riemann sphere without poles into the 
complex group $K^\bc$. (Clearly, the
loop circle $S^1$ is identified with the equator).  Then
 $$\k(l)(z)=l(e^{\ep k}z),\quad l\in LK^\bc,$$
 where $z$ is the usual complex coordinate on the Riemann sphere and $k$ is the same integer parameter which appears also in the standard WZW model. The maximally isotropic subgroup
 $G$ is nothing but $LK$ (the isotropy is a direct consequence of the fact that $(.\ve .)$ is real
 when restricted to $L\K$) and $B=L_+K^\bc$ is defined as the group of holomorphic maps from the Riemann sphere 
without the north pole
into the complex group $K^\bc$.
We require moreover, that the value  of this holomorphic map 
at the 
  south pole
 is an element of the group $AN$ defined by the Iwasawa decomposition $K^\bc=KAN$. 
 It is not difficult to establish the isotropy of $L_+K^\bc$ (cf. Sec 4.4.1 of \cite{K}). 
Finally, the existence of the global decomposition $D=LK^\bc=
\k(LK)(L_+K^\bc)=\k(G)B=BG$
was proved in \cite{PS}. 
 
 The symplectic structure on $D=LK^\bc$ is defined by the formula (6) . It turns out that the symplectic manifold $D$ has the structure of the (reduced) product $P\times P$ of two copies of a simpler symplectic
 manifold $P$ called chiral $q$-WZW model.  The points of the manifold $P$ are maps $m: \br \to K$,
 fulfilling the monodromy condition
 $$m(\si+2\pi)=m(\si)e^{-2\pi i x},$$
 where $x$  is an element of the Weyl alcove of the Cartan subalgebra of $Lie(K)$.
 The symplectic structure on $P$ is completely
characterized by the following matrix Poisson bracket:
$$  \{m(\si)\ptp m(\si') \bq =(m(\si)\otimes m(\si'))
 B_{\e}(x,\si-\si')+
 \e\hat r(\si-\si')(m(\si)\otimes m(\si')),\eqno(11)$$
where $\hat r(\si)$ is  a  trigonometric solution of the ordinary Yang-Baxter equation with spectral parameter
$$\hat r(\si)=r+C{\rm cotg}\jp\si,$$  $C$ is the Casimir element defined by
$$C=\sum_\mu H^\mu\ot H^\mu +\sum_{\al>0}{\vert\al\vert^2\over 2}
(E^{-\al}
\ot E^\al+E^{\al}\ot E^{-\al}). $$and  
$$r=\sum_{\al>0}{i\vert\al\vert^2\over 2}(E^{-\al}\ot E^\al-E^\al\ot
E^{-\al}).$$
The matrix $ B_{\e}(x,\si)$ is the so called Felder's elliptic r-matrix \cite{F} and it solves the dynamical
Yang-Baxter equations with spectral parameter. 
$$ B_{\e}(x,\si)=$$
$$ =-{i\over \k}\rho({i\si\over 2\k\e},{i\pi\over \k\e})
H^\mu\otimes H^\mu 
-{i\over \k}\sum_{\al}{\rn\over 2}\si_{\la 
\al,x\ra}
( {i\si\over 2\k\e},{i\pi\over \k\e})E^\al\otimes E^{-\al}.$$
 The elliptic functions $\rho(z,\tau),\si_w(z,\tau)$ are defined as  
(cf. \cite{F})
$$ \si_w(z,\tau)={\theta_1(w-z,\tau)\theta_1'(0,\tau)\over 
\theta_1(w,\tau)\theta_1(z,\tau)},
\quad \rho(z,\tau)={\theta_1'(z,\tau)\over \theta_1(z,\tau)}.$$
Note that  $\theta_1(z,\tau)$ is the Jacobi theta 
function\footnote{We have
$\theta_1(z,\tau)=\vartheta_1(\pi z,\tau)$ with 
$\vartheta_1$ in \cite{WHW}.}
$$ \theta_1(z,\tau)=-\sum_{j=-\infty}^{\infty}e^{\pi i (j+\jp)^2\tau 
+2\pi i(j+\jp)(z+\jp)},$$
 the prime ' means the derivative with respect to the
 first argument $z$ and  
the argument $\tau$ (the modular parameter )
is a nonzero complex number such that Im $\tau>0$.

\vskip1pc

\noindent We note that in the limit $\ep\to 0$ the symplectic structure (11) becomes that of the
ordinary chiral WZW model (cf. \cite{K}).  The time evolution in $q$-WZW model is the same
as in the non-deformed case, i.e.
$$[m(\si)](\tau)=m(\si-\tau),$$
where $\tau$ is the evolution parameter (time). 

\vskip1pc

\noindent The chiral model $P$ turns out to be  Poisson-Lie symmetric with respect to the natural
left action of $LK$ on $P$. For a detailed exposition
of this fact see \cite{K}, here we just mention what is the $\ep$-deformation of the standard
Kac-Moody commutation relation (10a). Thus we set
$$L(\si)=m(\si+i\k\ep)m^{-1}(\si-i\k\ep)$$
and we use the fundamental Poisson brackets (11) to find well-known relations of deformed current algebra \cite{RST}:
$$\{L(\si)\ptp L(\si')\}=(L(\si)\ot L(\si'))\e\hat r(\si-\si')
+\e\hat r(\si-\si')(L(\si)\ot L(\si')) $$  
$$  -(1\ot L(\si'))\e\hat r(\si-\si'+2i\e\k)(L(\si)\ot 1)
-(L(\si)\ot 1)\e\hat r(\si-\si'-2i\e\k)(1\ot L(\si')).\eqno(12)$$
\section{Concluding remarks}

Perhaps the most remarkable  result of our
 deformation program is the fact that the  structure of the deformed chiral WZW model
 is essentially characterized by the Felder elliptic dynamical r-matrix \cite{F}.
 In particular, we find quite amazing that from so small input (the group structure of $LK^\bc$)
 such a nice mathematical object like Felder's matrix naturally emerges.  
 We have shown another interesting thing, namely, the trigonometric deformed current algebra  (12) is   underlied by a more fundamental elliptic structure (11).

\end{document}